\documentclass[a4paper,11pt]{article}
\pdfoutput=1 

\usepackage{jcappub} 

\usepackage[T1]{fontenc} 
\usepackage{amssymb}
\usepackage{amsmath}
\usepackage{graphicx}
\usepackage{color}
\usepackage{pdflscape}

\newcommand{\cosmomc}{\texttt{CosmoMC}}
\newcommand{\camb}{\texttt{camb}}
\newcommand{\calAs}{\mathcal{A}_s}
\newcommand{\clamp}{10^9 A_s e^{-2\tau}}
\newcommand{\clampp}{A_s e^{-2\tau}}
\newcommand{\planck}{\textit{Planck}}
\newcommand{\plancklowT}{\planck\ lowT}
\newcommand{\wlowP}{WMAP9 lowP}
\newcommand{\whfilowP}{WMAP353 lowP}
\newcommand{\lfiwhfilowP}{WMAP/\planck\ lowP}
\newcommand{\plikTT}{\planck\ high TT}
\newcommand{\wmapTT}{WMAP9 high TT}
\newcommand{\lensing}{\planck\ lensing}
\newcommand{\priors}{high-$\ell$ priors}
\newcommand{\lowP}{\planck\ lowP}
\newcommand{\nquad}[1]{\hspace{#1em}}
\newcommand{\eightquad}{\nquad{8}}

\title{On the impact of large angle CMB polarization data on cosmological parameters}

\author[a,b,1]{Massimiliano Lattanzi,\note{Corresponding author.}}
\author[c,a,d]{Carlo Burigana,}
\author[e,f]{Martina Gerbino}
\author[c,d]{Alessandro Gruppuso}
\author[a,c]{Nazzareno Mandolesi}
\author[a,b]{Paolo Natoli}
\author[g,h]{Gianluca Polenta}
\author[i,l]{Laura Salvati}
\author[c,a,d]{Tiziana Trombetti}

\affiliation[a]{Dipartimento di Fisica e Scienze della Terra, Universit\`a di Ferrara, Via Giuseppe Saragat 1, I-44122 Ferrara, Italy}
\affiliation[b]{Istituto Nazionale di Fisica Nucleare, Sezione di Ferrara, Via Giuseppe Saragat 1, I-44122 Ferrara, Italy}
\affiliation[c]{INAF, Istituto di Astrofisica Spaziale e Fisica Cosmica di Bologna, Via Piero Gobetti 101, I-40129 Bologna, Italy}
\affiliation[d]{Istituto Nazionale di Fisica Nucleare, Sezione di Bologna, Via Irnerio 46, I-40126 Bologna, Italy}
\affiliation[e]{The Oskar Klein Centre for Cosmoparticle Physics, Department of Physics, Stockholm University, AlbaNova, SE-106 91 Stockholm, Sweden}
\affiliation[f]{The Nordic Institute for Theoretical Physics (NORDITA), Roslagstullsbacken 23, SE-106 91 Stockholm, Sweden}
\affiliation[g]{Agenzia Spaziale Italiana Science Data Center, Via del Politecnico snc, 00133, Roma, Italy}
\affiliation[h]{INAF - Osservatorio Astronomico di Roma, Via di Frascati 33, Monte Porzio Catone, Italy}
\affiliation[i]{Dipartimento di Fisica, Universit\`a La Sapienza, Piazzale Aldo Moro 2, I-00185 Roma, Italy}
\affiliation[l]{Istituto Nazionale di Fisica Nucleare, Sezione di Roma 1, Universit\`a di Roma Sapienza, Piazzale Aldo Moro 2, I-00185 Roma, Italy}

\emailAdd{lattanzi@fe.infn.it}

\abstract{
We study the impact of the large-angle CMB polarization datasets publicly released by the WMAP and \planck\ satellites on the estimation of cosmological parameters of the $\Lambda$CDM model. To complement large-angle polarization, we consider the high resolution (or ``high-$\ell$'') CMB datasets from either WMAP or \planck\, as well as CMB lensing as traced by \planck's measured four point correlation function. In the case of WMAP, we compute the large-angle polarization likelihood starting over from low resolution frequency maps and their covariance matrices, and perform our own foreground mitigation technique, which includes as a possible alternative \planck\ 353 GHz data to trace polarized dust. We find that the latter choice induces a downward shift in the optical depth $\tau$, roughly of order $2\sigma$, robust to the choice of the complementary high resolution dataset. When the \planck\ 353 GHz is consistently used to minimize polarized dust emission, WMAP and \planck\ 70 GHz large-angle polarization data are in remarkable agreement: by combining them we find  $\tau = 0.066 ^{+0.012}_{-0.013}$, again very stable against the particular choice for high-$\ell$ data. We find that the amplitude of primordial fluctuations $A_s$, notoriously degenerate with $\tau$, is the parameter second most affected by the assumptions on polarized dust removal, but the other parameters are also affected, typically between $0.5$ and $1\sigma$. In particular, cleaning dust with \planck's 353 GHz data imposes a $1\sigma$ downward shift in the value of the Hubble constant $H_0$, significantly contributing to the tension reported between CMB based and direct measurements of the present expansion rate. On the other hand, we find that the appearance of the so-called low $\ell$ anomaly, a well-known tension between the high- and low-resolution CMB anisotropy amplitude, is not significantly affected by the details of large-angle polarization, or by the particular high-$\ell$ dataset employed. 
}

\begin{document}
\maketitle
\flushbottom

\section{Introduction}
\label{intro}

The cosmic microwave background (CMB) anisotropies, tiny temperature fluctuations in the primordial blackbody radiation averaging at 2.72548 $\pm$ 0.00057 K \citep{Fixsen:2009ug}, are one of the landmark pieces of evidence of the Big Bang theory. The wealth of information encoded in the CMB has significantly helped cosmology to turn into a precision science, and allowed us to describe cosmic evolution starting from its very primordial stages. The CMB anisotropies are linearly polarized \citep{Hu:1997hv}, the polarization pattern being generated in two distinct cosmological epochs. The first one is the recombination era ($z \simeq 1100$) during which, as the Universe expands and cools, the diffusion length of photons in an increasingly neutral medium became large enough to reveal
the quadrupole moment of the local anisotropy pattern at the so-called last scattering surface. The Universe then became almost neutral and transparent to radiation, since Thomson scattering, the main physical process involved, 
was no longer efficient in coupling matter and radiation. Much later, during the early phases of galaxy and star formation,
photoionizing radiation able to escape from the bound structures was injected in the intergalactic medium. As a consequence, the medium  became ionized again (hence the term reionization) providing free electrons to
interact through Thomson scattering with CMB photons once more. 
We know that at $z \gtrsim 6$, the intergalactic medium was already almost fully ionized, as shown by the change with redshift of the absorption spectra of distant quasars or gamma ray-burtsts 
on the blue side of the Lyman-$\alpha$ emission \citep{Gunn:1965hd,Fan:2005es,Bolton:2011vb,Chornock:2013una,McGreer:2014qwa}.
This process modifies the pre-existing CMB polarization pattern, which is seeded again from 
the local anisotropy quadrupole. Owing to the size of the Hubble horizon at reionization, though, this new contribution is only seen at large angular scales, as a characteristic bump at low multipoles ($\ell \lesssim 10$) in the pure polarization CMB spectra, as well as in the temperature to polarization correlation.

The onset of the reionization epoch mainly
determines the Thomson optical depth $\tau$ between the observer and the CMB last scattering surface,
one of the six cosmological parameters of the standard $\Lambda$CDM model \cite{Ade:2015xua} (see also Refs. \cite{Mortonson:2007hq,Pandolfi:2011kz,Trombetti:2012,Douspis:2015nca,Oldengott:2016yjc,Heinrich:2016ojb,Adam:2016hgk} for analyses going beyond the simple $\tau$ parametrisation). 
All primary CMB angular power spectra are affected by reionization in having their amplitude suppressed by a factor $e^{-2\tau}$. Such a suppression, which equally affects intensity and polarization spectra, arises due to re-scattering of photons by free electrons and is effective at all scales but the largest ones that were outside the horizon at reionization; hence the ``bump'' at low $\ell$'s. This loss in anisotropy power at non-zero $\tau$ is clearly degenerate with the amplitude of primordial scalar fluctuations $A_s$ which source CMB anisotropy.  On the other hand, only polarization spectra inherit the characteristic low $\ell$ increase, or bump, mentioned above, whose amplitude is $\propto \tau^2$. This unique signature is exploited to disentangle $\tau$ and $A_s$ and better infer the value of both these $\Lambda$CDM model parameters. 

The first estimate of $\tau$ from CMB measurements was provided by the Wilkinson Microwave Anisotropy Probe (WMAP). In the nine-year legacy results, the WMAP collaboration reports $\tau$ = 0.089 $\pm$ 0.014 \cite{Hinshaw:2012aka}. \planck\ \cite{Adam:2015rua} has shown \citep{Ade:2013kta,Aghanim:2015xee,Ade:2015xua} that the WMAP estimate is likely biased high due to residual dust polarization, imperfectly accounted for by the simple dust model the WMAP team employed, since they lacked of tracers based on direct observations. \planck's 353 GHz channel, on the other hand, is an ideal monitor of dust polarization. Considering still WMAP large-scale polarization, but using the 353 GHz channel to mitigate dust, $\tau$ shifts down to $0.075\pm 0.013$ \citep{Ade:2013kta}. The 2015 analysis of \planck\ reports $\tau = 0.067\pm0.023$ using only \planck\ low-resolution data (from the 70 GHz channel of its low frequency instrument) and priors on cosmological parameters fixed by small-scale CMB information \citep{Aghanim:2015xee}.
More recently, the Planck collaboration has published a new estimate of $\tau = 0.055\pm0.009$ \citep{Aghanim:2016yuo} using only low-$\ell$ polarization $EE$ cross-spectra built from the 100 and 143 GHz channels of the high frequency instrument\footnote{The new products, and the associated analysis pipelines, 
presented in Ref. \cite{Aghanim:2016yuo} are referred to as ``pre-2016''.}, unlike the \planck\ 2015 low-$\ell$ results that also make use of TT and TE information. This new value is fully compatible with the previous one from \planck\ and displays 
a smaller uncertainty. It is also quite stable with respect to the choice of the likelihood method and of the cross-spectra estimator, providing
consistent results albeit, in some cases, with larger error bars. Remarkably enough, Ref. \cite{Aghanim:2016yuo} also shows how building cross-spectra
from the 70 GHz channel of \planck's low frequency instrument and either the 100 or 143 GHz channel, still gives consistent results: for example,
the 70x143 cross spectra give $\tau = 0.053^{+0.012}_{-0.016}$. Implications of the pre-2016 data for cosmic reionization have been investigated 
by the Planck collaboration in Ref. \cite{Adam:2016hgk}, considering various parameterisations of the reionization history. For the standard assumption of instantaneous reionization, it is found, using a different likelihood code, that $\tau = 0.058\pm 0.012$, thus with a 30\% larger uncertainty with
respect to the value quoted in Ref. \cite{Aghanim:2016yuo} under the same assumption.
Unlike the results previously reported in the 2015 \planck\ parameters paper \cite{Ade:2015xua}, though, the  latest \planck\ $\tau$ measurements reported in Refs. \cite{Aghanim:2016yuo} and \cite{Adam:2016hgk} are not accompanied by a public CMB likelihood module, which will be made available as part of the \planck\ legacy release, in the near future.
We do not consider the pre-2016 results any further in this paper.

The aim of the present work is to study how the choice of existing low-multipole polarization data impacts cosmological parameter estimates,
by looking at different combinations of the public low- and high-$\ell$ datasets provided by WMAP9\footnote{We acknowledge 
the use of the products available at the Legacy Archive for Microwave Background Data Analysis, LAMBDA (\url{http://lambda.gsfc.nasa.gov/}).} and \planck\footnote{We acknowledge 
the use of the products available at the Planck Legacy Archive (\url{http://www.cosmos.esa.int/web/planck/pla}).}. In particular, we discuss the impact of using the \planck\ 353 GHz channel as a dust tracer to also clean the large-scale WMAP9 polarization.

While it has been known since the \planck\ 2013 release  that the  \planck\ 353 GHz channel induces a significant ($\gtrsim 1\sigma$) shift in WMAP's value of the optical depth $\tau$ \citep{Ade:2013kta} and 
brings the WMAP9 and \planck\ estimates of this parameter to very good agreement \citep{Aghanim:2015xee}, a systematic study of the impact of the public low $\ell$ polarization datasets on cosmological parameters, in view of 
the particular choice of high-$\ell$ data employed (that is, either \planck\ or WMAP9), was not available in the literature. This paper aims at filling this gap.
We also study the impact of the dust cleaning procedure on the so-called ``low-$\ell$ anomaly'', i.e. the tension between the values of the amplitude of primordial perturbations estimated from the low- ($\ell<30$) and high- ($\ell \ge 30$) ell portions of the TT power spectrum. The low-ell anomaly has been found to 
drive the differences in parameter values obtained from distinct subsets of the high-$\ell$ data \citep{Addison:2015wyg,Aghanim:2016sns}. Even 
though the significance of these shifts is low, once ``look-elsewhere'' effects are taken into account \citep{Aghanim:2016sns}, it is nevertheless interesting to assess how much the low-$\ell$ anomaly that causes them depends on the choice of low-$\ell$ polarization data. By comparing parameter estimates obtained from the low- and high-$\ell$ datasets, we test the robustness of the anomaly to the low resolution cleaning 
procedure for polarized dust, as well as of the particular choice of the high-$\ell$ datasets.

In our analysis we focus on the standard $\Lambda$CDM model and its parameters, and thus do not explore, among others, the 
link between the low-$\ell$ polarization data and a non-standard lensing amplitude (caused either by modifications to gravity, or introduced
as a proxy parameter for systematics); for such an analysis, see e.g. \cite{Couchot:2015eea}.

The paper is organized as follows. In Sec. \ref{dataset} we describe the datasets employed throughout the analysis and the corresponding likelihood modules. In Section \ref{results} we present our results, and in Sec. \ref{conclusions} we draw our conclusions.

\section{Method and data sets}
\label{dataset}

\subsection{Description of the low-$\ell$ data}
\label{lowdataset}

At low multipoles ($\ell < 30$), we consider the standard joint temperature-polarization, pixel-based machinery for the likelihood,
as done in the 2015 \planck\ release \citep{Aghanim:2015xee}. Thus, our low-$\ell$ datasets always consist of low resolution ($N_\mathrm{side}=16$ in Healpix\footnote{\url{http://healpix.jpl.nasa.gov/}; see also \citep{Gorski:2004by}.}) 
maps of temperature and polarization anisotropies, and of the related noise covariance matrices. Note that the publicly released WMAP9 likelihood code relies instead on $N_\mathrm{side}=8$ polarization maps: our choice of $N_\mathrm{side}=16$ is primarily motivated by the need to standardise WMAP9 and \planck\ low resolution products.
 
As low-$\ell$ temperature data, we always employ the \planck\ \texttt{Commander} CMB map, obtained combining the \planck\ observations between 30 and 857 GHz 
\cite{Adam:2015rua}, the 9-year WMAP observations between 23 and 94 GHz \citep{Bennett:2012zja}, and the 408 MHz sky maps from \citet{Haslam:1982zz}, as described in Refs. \cite{Adam:2015tpy,Adam:2015wua,Aghanim:2015xee}
(see also Refs. \cite{Eriksen:2004ss,Eriksen:2007mx} for a description of the \texttt{Commander}  component-separation algorithm), 
The \texttt{Commander} map covers 94\% of the sky and will hereafter be denoted as ``\plancklowT''. 
 
For  low-$\ell$ polarization, we consider three different datasets, all based on WMAP9 data; two of them 
are augmented by also considering information from \planck\ low-resolution polarization maps. 

(1) For the dataset denoted ``\whfilowP'', we consider the WMAP9 raw low resolution frequency maps 
for the Ka, Q, and V channels and the related noise covariance matrices that are publicly available. 
The WMAP team has decided not to include their W band data for low $\ell$ polarization analysis, on the grounds of evidence of systematic contamination \citep{Bennett:2012zja}. Our re-analysis of WMAP's W channel confirms the presence of excess power of likely spurious origin, so we consistently disregard these data in what follows.

To minimize the impact of polarized foregrounds, we implement a template fitting procedure that relies on the following scheme. Each linear polarization map 
$\bf {m}_{\nu}=\big[\bf{Q}_{\nu},\bf{U}_{\nu}\big]$, ($\bf {Q}_{\nu},\bf {U}_{\nu}$ being the Stokes parameters maps at frequency band $\nu$) is
masked by the WMAP9 processing mask \citep{Bennett:2012zja} and used
to find two scaling coefficients $\alpha_{\nu}$ and $\beta_{\nu}$, assumed constant over the available sky, that minimise 
the form $\chi^2_\mathrm{S+N} = \mathbf{\tilde m}_{\nu}^\mathrm{T} \bf{C}^{-1}_{\nu} \tilde{{\bf m}}_{\nu}$,
with $\tilde{{\bf m}}_{\nu} = {\bf m}_{\nu} - \alpha_{\nu} {\bf t}_{\mbox {s}} - \beta_{\nu} {\bf t}_{\mbox {d}} $, 
 ${\bf t}_{\mbox {s}}$ and ${\bf t}_{\mbox {d}}$ being the templates for synchrotron and dust emissions, chosen 
as the WMAP9 K-band and the \planck\ 353 GHz maps respectively.
The covariance matrix ${\bf C}_{\nu}= {\bf S}+ {\bf N}_{\nu}$ is taken as the sum of the signal\footnote{We use thermodynamic temperature units so each channel has the same signal covariance matrix.},
${\bf S}$, and noise, ${\bf N}_{\nu}$, matrices, and the signal matrix is built through  harmonic expansion \citep{Tegmark:2001zv} assuming the WMAP 9-year fiducial model;  
we have nonetheless verified that using in its place the \planck\ 2015 fiducial induces negligible differences in the estimated scaling coefficients. The WMAP team \citep{Bennett:2012zja} has used a similar procedure to derive their scaling coefficients, although neglecting the CMB signal in the $\bf{C}_\nu$ matrices. We have verified that the latter approximation induces only differences that lie well within the error budget. We nonetheless also report (see table \ref{tabellauno} below) the $\chi^2_\mathrm{N}$ figures we find assuming $\bf{C}_\nu \equiv \bf{N}_\nu$ only. 

Once $\alpha_{\nu}$ and $\beta_{\nu}$ are found, the foreground-reduced polarization maps $\hat{{\bf m}}_{\nu} $ are obtained as:
\begin{equation}
\hat{{\bf m}}_{\nu} = ( {\bf m}_{\nu} - \alpha_{\nu} {\bf t}_{\mbox {s}} - \beta_{\nu} {\bf t}_{\mbox {d}} ) / (1-  \alpha_{\nu} - \beta_{\nu})
\, ,
\label{foreredmap}
\end{equation}
and the associated noise covariances $\hat{{\bf N}}_{\nu}$ are: 
\begin{equation}
\hat{{\bf N}}_{\nu} = \left( {\bf N}_{\nu}  + \delta {\alpha}^2 _{\nu} \, {\bf F}_{\mbox {s}} + \delta {\beta}^2 _{\nu} \, {\bf F}_{\mbox {d}} \right) / (1-{\alpha}_{\nu}-\beta_{\nu})^2
\, ,
\label{foreredcov}
\end{equation}
where $\delta {\alpha}$ and  $\delta {\beta}$ are the standard errors associated with the estimated scalings, and 
${\bf F}_{\mbox {s,d}}\equiv  \mathbf{t_{s,d}^{\phantom{\mathrm{T}}}} \mathbf{t_{s,d}^\mathrm{T}}$
are the covariance contributions for synchrotron and dust emission respectively, built from the outer products of the corresponding templates. In table \ref{tabellauno} we list 
the values we find for the scaling coefficients assuming the processing mask, and the corresponding $\chi^2$ figures. Fixing the scaling coefficients just found, we also report $\chi^2$ values outside of WMAP's cosmological P06 mask (see Ref. \cite{Page:2006hz} for a description of WMAP polarization masks), which leaves about 73\% of the sky. The reduced $\chi^2$ values we find are consistent with unity for all bands, in all masks considered. From the estimated scaling coefficients and their associated standard errors we compute our best fist estimate for the synchrotron spectral index as $-3.152 \pm 0.019$ and for the dust spectral index as $1.332 \pm 0.024$ ($1\sigma$ errors). These estimates were fitted considering the scaling coefficients derived for K$_a$, $Q$ and $V$ and their $1\sigma$ errors, while only small regularization errors were associated with both foreground templates (whose scaling coefficients are fixed to unity). 
We find that our estimates are robust to the fitting procedure employed, as well as to our assumptions for errors in the templates.
In particular, adding the calibration errors quoted for the WMAP9 and \planck\ 353 GHz frequency maps to standard errors estimated for the scaling coefficients, we find
a synchrotron spectral index of $-3.154 \pm 0.022$ and a dust spectral index of $1.332 \pm 0.025$. Furthermore, we checked that these results do not change significantly including WMAP W channel.
The values we find for the spectral indexes are in agreement with other measurements, including those reported in Ref. \cite{Aghanim:2015xee}, where the \planck\ 353 GHz channel was also used as a dust tracer, but otherwise only the \planck\  70 GHz and 30 GHz channels were considered. Remarkably, the synchrotron spectral index is also in agreement at $1\sigma$ with the WMAP9 result derived by the WMAP team applying the template 
fitting method to polarization maps. On the contrary, they found a steeper dust spectrum, with an index ranging from 1.41 to 1.5. Consequently, we repeated for comparison our fit but considering only the rescaling coefficients 
in the WMAP channels specified above and including calibration uncertainties.
For the synchrotron spectral index we find very similar results ($-3.152 \pm 0.022$). For the dust spectral index, on the other hand, we find $1.420 \pm 0.091$, in agreement with WMAP team results. 
Clearly, our results for the dust including the \planck\ 353 GHz frequency map are driven by highest frequencies. 
It is interesting to note that a gray body spectrum with dust grain temperature $T_{dg} = 19.6\,$K and grain emissivity index $\beta_{dg} = 1.59$, as found in Ref. \cite{Adam:2014bub} at intermediate and high Galactic latitudes,
predicts an effective spectral index of 1.33 between 93 GHz and 353 GHz and of 1.52 between 33 GHz and 93 GHz, in agreement with the results described above.
The spectral indices derived with our template fitting procedure applied to WMAP and \planck\ 353 GHz polarization maps are then in line with those quoted in the recent literature
and, in particular, provide an evidence of the gray body spectrum of Galactic thermal dust emission.

\begin{table}
\centering
\label{tabellauno}
\begin{tabular}{|cccccccc|}
\hline
$\nu$ & $ \alpha $ & $ \beta $ & $N_\mathrm{pix}$ & $\chi_\mathrm{S+N}^2/ N_\mathrm{pix}$ & $\chi_\mathrm{N}^2/ N_\mathrm{pix}$ & $\tilde\chi^2_\mathrm{S+N}/ \mathcal{N}_\mathrm{pix}$  & $\tilde\chi^2_\mathrm{N}/ \mathcal{N}_\mathrm{pix}$\\
\hline
Ka & $0.315\pm 0.003$ & $0.0031\pm 0.0004$ & 5720 & 1.026 & 1.045 & 1.012 & 1.029\\
Q & $0.163\pm 0.003$ & $0.0039\pm 0.0004$ & 5742 & 0.9830 & 1.004& 0.9778 & 0.995\\
V & $0.047\pm 0.003$ & $0.0076\pm 0.0004$ &  5894 & 0.9627  & 0.977 & 0.9606 & 0.974\\
\hline
\end{tabular}
\caption{Best-fit scaling coefficients and $\chi^2$ values for WMAP9 low $\ell$ polarization frequencies.
Scalings are obtained using the processing mask, see text as well as Ref. \cite{Bennett:2012zja}. The $\chi^2_\mathrm{S+N}$ values reported are computed as described in the text, outside of the processing mask of each channel, leaving $N_\mathrm{pix}$ good pixels in the $[Q,U]$ maps. We also report $\chi^2_\mathrm{N}$ values, assuming only the noise covariance matrix for each channel (more similarly to WMAP team's analysys). Differences between these two approaches are weak. The last columns report $\tilde\chi^2$ values, for both S+N and N, computed using the same method as above, but outside of the P06 mask used for parameter estimation, leaving $\mathcal{N}_\mathrm{pix} = 4510$ good pixels. To compute these last $\tilde\chi^2$ values the scaling coefficients derived in the processing mask have been assumed, estimated separately for S+N and N: the scalings quoted in the first two columns are for the former case, though the latter produces figures that are well within errors.}
\end{table}

We finally proceed to create our final polarization maps by performing a noise weighted average of the Ka, Q and V channels: 
\begin{equation}
{\bf m}_{Ka+Q+V} = {\bf N}_{Ka+Q+V} ({\hat {\bf N}}_{Ka}^{-1} \hat{{\bf m}}_{Ka} +{\hat {\bf N}}_Q^{-1} \hat{{\bf m}}_{Q} + {\hat {\bf N}}_V^{-1} \hat{{\bf m}}_{V} )
\, ,
\label{KaQV}
\end{equation}
where
${\bf N}_{Ka+Q+V}^{-1} = {\hat {\bf N}}_{Ka}^{-1} +{\hat {\bf N}}_Q^{-1} +{\hat {\bf N}}_V^{-1}$.

(2) As our second dataset, we adopt the WMAP9 low-$\ell$ data set in polarization, where we use as a dust tracer 
the model employed in WMAP9 own analysis \citep{Page:2006hz}, 
in place of the \planck\ 353 GHz map. This dataset is similar to the one employed by the WMAP team for their likelihood analysis, except that our version has resolution of $N_\mathrm{side}=16$. We nonetheless use the scalings $\alpha_{\nu}$ and $\beta_{\nu}$ reported in Ref. \cite{Bennett:2012zja}, after having verified that our own estimate reproduces their results within errors.
Since this dust model does not contain any CMB signal, the foreground-reduced maps
and corresponding covariances are obtained as in Eqs.~(\ref{foreredmap}) and (\ref{foreredcov})
but with a rescaling factor given by $1/(1-  \alpha_{\nu})$ or $1/(1-  \alpha_{\nu} )^2$ respectively.
The dataset based on noise weighted co-addition [again following Eq.~(\ref{KaQV})] of the Ka, Q and V foreground-reduced maps obtained in this way is referred hereafter as 
``\wlowP''.

(3) For the third dataset considered in this article, denoted ``\lfiwhfilowP'', we use the same cleaning procedure 
as the \whfilowP\ dataset, \emph{i.e.}, we use the \planck\ HFI 353 GHz map as a dust template, but we add to the WMAP Ka, Q and V maps the publicly available \planck\ LFI 70 GHz map \citep{Aghanim:2015xee}.
The combination of these maps is performed, similarly to Eq.~(\ref{KaQV}), 
through a standard noise-weighted average over the union of the useful fraction of the sky observed by each experiment.
Pixels observed by just one experiment are simply added to the dataset; this formally amounts to 
a weighted average with infinite noise for the missing experiment.

The computation of the likelihood associated with these three datasets is performed following the 
standard pixel-based approach described in Ref. \cite{Tegmark:2001zv}. In particular,
we compute the exact joint likelihood $\mathcal{L}(\mathbf{m}|C_\ell)$ of the 
temperature and polarization maps given the theoretical model (up to $\ell_\mathrm{max} = 29$), using
\begin{equation}
\mathcal{L}(\mathbf{m}|C_\ell) = \frac{1}{2\pi \left| \mathbf{C} \right|^{1/2}} \exp\Bigg(-\frac{1}{2} \mathbf{m}^\mathrm{T}\mathbf{C}^{-1} \mathbf{m}\Bigg),\,
\end{equation}
where $\mathbf{m} = [\mathbf{T},\,\mathbf{Q},\,\mathbf{U}]$ is the TQU map, and $\mathbf{C}$ is the 
corresponding covariance matrix, containing signal and noise contributions, 
that takes into account
the full covariance structure of the data (including temperature-polarization correlations). The signal part of the covariance matrix
is built using multipoles up to $\ell_\mathrm{max} = 47$; multipoles in the range $30 \le \ell \le 47$ are fixed to a fiducial model.
The temperature map - always the \planck\ \texttt{Commander} map - is masked with the \planck\ 2015 temperature mask,
that uses 94\% of the sky. The polarization maps described above are masked using WMAP's P06 mask (for the \wlowP\ and \whfilowP\ datasets),
that keeps 73\% of the sky, or the intersection of the P06 and \planck's R1.50 cosmological mask (for the \lfiwhfilowP dataset), keeping 74\% of the sky 
\citep{Aghanim:2015xee}.
In the following, since the temperature dataset at low multipoles is always the same,
we shall often omit to mention it when reporting our result, and identify the full low-$\ell$ dataset
only through the polarization part of the dataset itself. 
Thus, when reporting results obtained, for example,
from \whfilowP, it should be understood that we are actually referring to the joint \plancklowT\ + \whfilowP\
dataset.
While we do not explicitly make use of the \planck\ official low-resolution dataset by itself, the analysis 
reported in Ref. \cite{Aghanim:2015xee} has been performed with a likelihood machinery similar to the one employed here.
Thus, results for \planck\ 70 GHz as quoted from the Planck collaboration papers can be directly comparable to the estimates presented in this work.

\subsection{Description of the high-$\ell$ data}

At high multipoles (\emph{i.e.}, $\ell \ge 30$), we consider the temperature power spectra from WMAP9 and \planck\, as well as
the power spectrum of the lensing potential derived from the \planck\ data. The WMAP9 TT power spectrum
\citep{Bennett:2012zja}
is based on V-band and W-band cross-power-spectra, and is computed using Gibbs sampling for $2 \le \ell \le 32$, and an optimal unbiased
quadratic estimator in the range $ 32 < \ell \le 1200$. Similarly, the \planck\ TT power spectrum in the range $30 \le \ell \le 2500$ is derived from an approach based on pseudo-$C_\ell$
estimators, using \planck\ HFI data (in particular, data from the 100 GHz, 143 GHz and 217 GHz channels, see Ref. \cite{Aghanim:2015xee}).
We refer to the two datasets consisting of the WMAP9 ($30 \le \ell \le 1200$) and \planck\ ($30 \le \ell \le 2500$) TT power spectra as
``\wmapTT'' and ``\plikTT'', respectively.
The likelihood functions associated with these datasets are computed using the public routines provided by the respective
collaborations.

We also consider the information encoded in the 4-point correlation function of the temperature anisotropies as measured by \planck,
that allows us to reconstruct the lensing potential along the line of sight between the observer and the last scattering surface. This dataset
and the corresponding publicly released likelihood, which we shall denote with ``\lensing'',  are described in detail in Ref. \cite{Ade:2015zua}.

A complete list of the datasets used in our analysis, together with a short description of each, can be found in table \ref{tab:datasets}.

\subsection{Method \label{method}}

\begin{table}
\centering
\scriptsize
\begin{tabular}{|ll|}
\hline
Name & Description \\
\hline
\plancklowT & \planck\ Commander temperature map ($\ell <30 $). \\[0.2cm]
\wlowP & WMAP9 polarization maps dust-cleaned using WMAP team's model ($\ell<30$). \\[0.2cm]
\whfilowP & WMAP9 polarization maps dust-cleaned using the \planck\ 353 GHz map ($\ell<30$). \\[0.2cm]
\lowP & \planck\ polarization maps dust-cleaned using the \planck\ 353 GHz map ($\ell<30$). \\[0.2cm]
\lfiwhfilowP & Noise-weighted WMAP and \planck\ polarization maps (353-cleaned, $\ell<30$).\\[0.2cm]
\plikTT & \planck\ high-$\ell$ TT likelihood ($30 \le \ell \le 2500$). \\[0.2cm]
\wmapTT &  WMAP9\ high-$\ell$ TT likelihood ($30 \le \ell \le 1200$). \\[0.2cm]
\lensing & \planck\ lensing likelihood ($40 \le \ell \le 400$)  \\[0.2cm]
\priors & Gaussian priors derived from a joint likelihood using \plancklowT, \lowP\ and \plikTT: \\[0.1cm]
& $\omega_b = 0.022222 \pm 0.00023,\ \omega_c = 0.1197 \pm 0.0022,\ 100\theta = 1.04085 \pm 0.00047, n_s = 0.9655 \pm  0.0062$.  \\
\hline
\end{tabular}
\caption{Abbreviations for the several datasets (and combinations thereof) employed throughout this paper.} \label{tab:datasets}
\end{table}

We use \cosmomc\ \citep{Lewis:2002ah}, interfaced with the Boltzmann code \camb\ \citep{Lewis:1999bs}, as a Monte Carlo engine to derive constraints on the cosmological
parameters of the standard $\Lambda$CDM model, namely: the baryon and cold dark matter densities $\omega_b \equiv \Omega_b h^2$
and $\omega_c \equiv \Omega_c h^2$, the angle $\theta$ subtended by the sound horizon at recombination, the reionization optical depth 
$\tau$, the logarithmic amplitude of the primordial scalar perturbations $\ln\left(10^{10} A_s\right)$ and their spectral index $n_s$. 
We marginalize over a number of nuisance parameters describing residual astrophysical foregrounds and various instrumental uncertainties 
(e.g. calibration uncertainties); see \citet{Aghanim:2015xee} and \citet{Bennett:2012zja}. 

We derive parameter estimates from various combinations of the \planck\ and WMAP9 datasets, by interfacing
the relevant likelihood functions to \cosmomc.
We start by considering the three low-$\ell$ polarization datasets described above (\wlowP,\ \whfilowP,\ \lfiwhfilowP), always complemented by \plancklowT, with Gaussian priors on $\omega_b$, $\omega_c$, $\theta$ and $n_s$,
that are only poorly (or not at all) constrained by large-scale observations. The Gaussian priors on these
``high-$\ell$'' parameters are based on the
estimates from analysis of the \plancklowT,\ \lowP, \plikTT\ data presented in \citet{Ade:2015xua}, and
are reported in our table \ref{tab:datasets}.
Then we combine each of the low-$\ell$ datasets 
with either the \wmapTT\ or the \plikTT\ datasets, for a total of six baseline low+high-$\ell$ datasets. Each of these is
considered either by itself or in combination with the \planck\ lensing likelihood.
Moreover, in order to better assess the weight of each kind of CMB dataset (low-$\ell$, high-$\ell$, lensing) in constraining
the parameters of interest, we also study the two combinations \wmapTT + \lensing\ and \plikTT + \lensing\, \emph{i.e.}, 
without including any low-$\ell$ information. In addition to the results from parameter runs performed 
for the present analysis, we shall also, for comparison, refer to constraints that were derived by the Planck collaboration and reported in the \planck\ 2015 likelihood \citep{Aghanim:2015xee} and
parameters \citep{Ade:2015xua} papers.

\section{Results}
\label{results}

\begin{figure*}
\begin{center}
\includegraphics[width=0.95\linewidth,keepaspectratio]{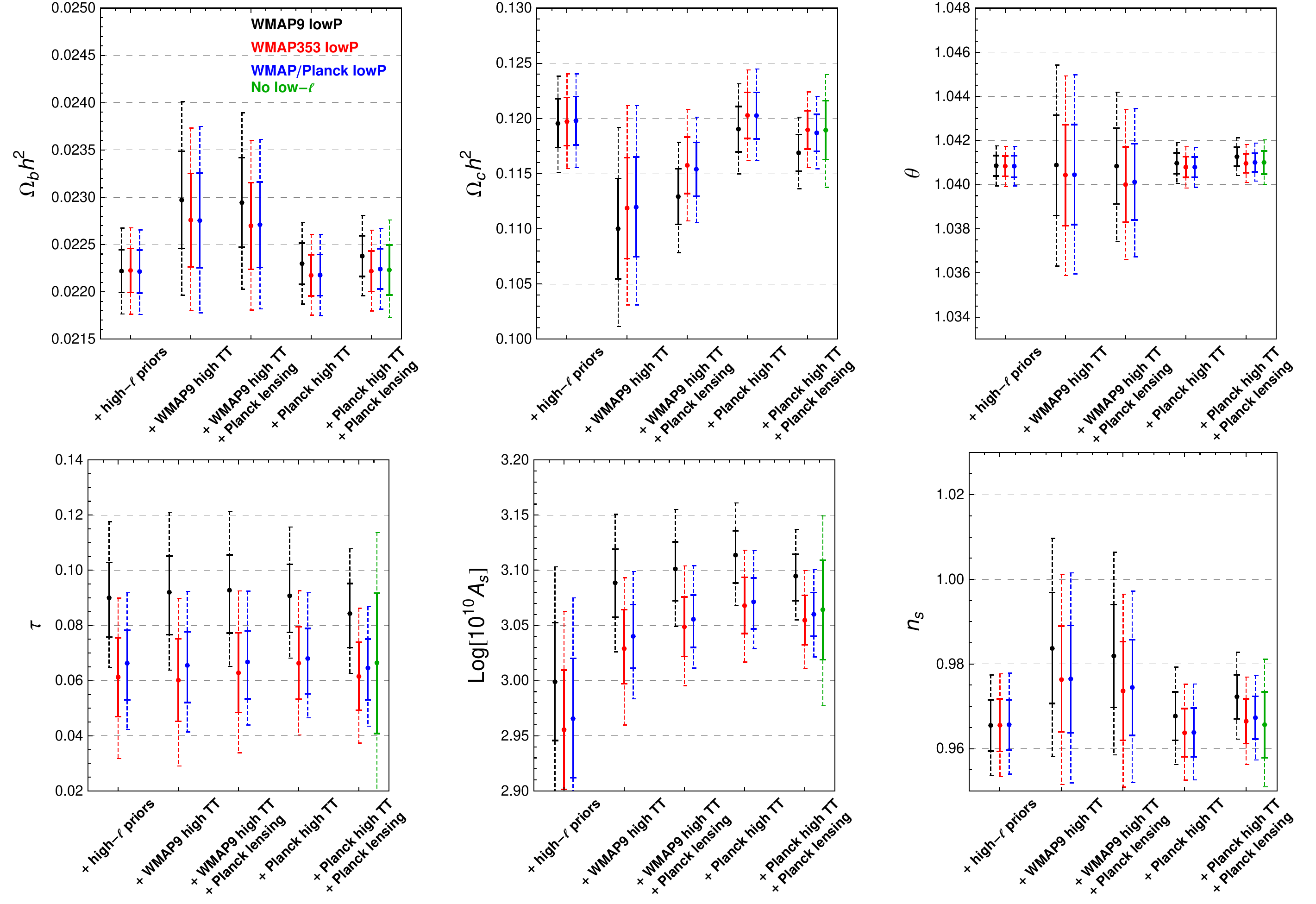}%
\caption{\label{fig:whisk_base} 68\% (solid) and 95\% (dashed) confidence intervals for the base $\Lambda$CDM parameters from various dataset combinations.
The color corresponds to the low-$\ell$ dataset, as indicated in the legend. From left to right, the groups of whiskers correspond to the following high-$\ell$ dataset combinations: i) \priors, ii) \wmapTT, iii) \wmapTT\ + \lensing, iv) \plikTT,  v) \plikTT\ + \lensing.}
\end{center}
\end{figure*}

\begin{figure*}
\begin{center}
\includegraphics[width=0.95\linewidth,keepaspectratio]{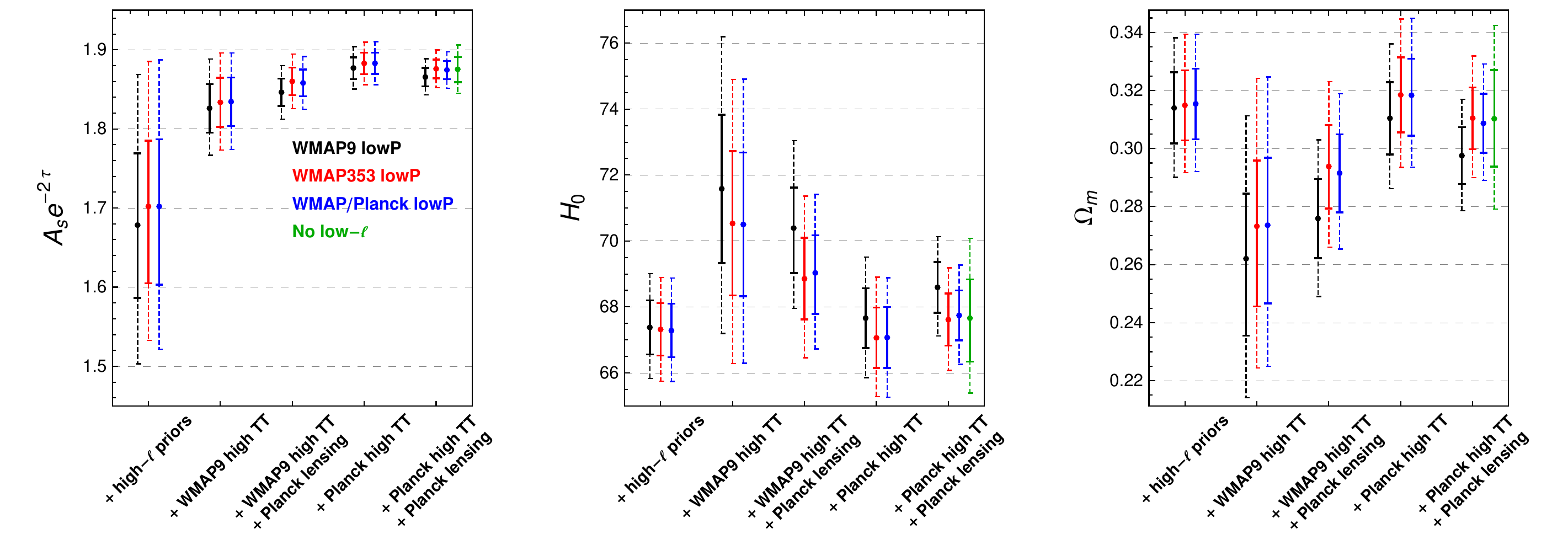}%
\caption{\label{fig:whisk_derived} The same as figure \ref{fig:whisk_base}, for derived parameters of interest. }
\end{center}
\end{figure*}

In the whiskers plots in Figs. \ref{fig:whisk_base} and \ref{fig:whisk_derived} we show the estimates of the six base $\Lambda$CDM and of selected derived parameters, obtained from the dataset combinations described above. There are several interesting trends in this plot that deserve to be highlighted; let us start by discussing the optical depth parameter $\tau$. It is clear from the relevant panel of figure \ref{fig:whisk_base} that the WMAP9 large-scale polarization data,
when cleaned using the WMAP dust model (black whiskers), yield a systematically higher value of $\tau$ (at the $\sim 2 \sigma$ level) with respect to the same data cleaned with 
the \planck\ 353 GHz map as a dust template (red whiskers). For example, from the low-$\ell$ data combined with Gaussian priors on the ``high-$\ell$'' parameters we get\footnote{We recall that here and in the following, it should always be understood that the ``lowP'' datasets are complemented by the \plancklowT\ data.}
\begin{subequations}
\begin{align}
&\tau = 0.090^{+0.013} _{-0.014}, \quad \textrm{\wlowP\ + \priors}, \label{eq:tau_WP9}\\
&\tau = 0.061\pm 0.014, \quad \textrm{\whfilowP\ + \priors} \label{eq:tau_WP353},
\end{align}
that is, the inclusion of the \planck\ 353 GHz polarized dust template immediately induces a downward $2\sigma$ shift
in $\tau$.
This value from \whfilowP\ is in very good agreement with the value derived using only the \planck\ 70 GHz for the low-$\ell$ polarization, 
reported in \citet{Aghanim:2015xee}:
\begin{align}
&\tau = 0.067\pm0.023, \quad \textrm{\lowP\ + \priors}\, \label{eq:tau_lowP} .
\end{align}
In fact, as also reported in Ref. \cite{Aghanim:2015xee}, the half-difference map between the \planck\ 70GHz and WMAP polarization
maps, when both are cleaned with the \planck\ 353GHz polarized dust template, is consistent with a noise-only, null-signal map.
Given their consistency, the \planck\ and WMAP low-ell polarization datasets can be combined by taking their noise-weighted sum, to form a joint set 
with higher signal-to-noise, as discussed in the previous section. From this joint dataset we get
\begin{align}
&\tau = 0.066^{+0.012}_{-0.013}, \quad \textrm{\lfiwhfilowP\ + \priors} \label{eq:tau_WPLFI}\,
\end{align}
\end{subequations}
\emph{i.e.}, a value that is very consistent with both  (\ref{eq:tau_WP353}) and (\ref{eq:tau_lowP}).
The uncertainty in the determination of $\tau$ is always roughly the same for all datasets using WMAP at low-$\ell$'s, implying
that the use of the \planck\ 353 GHz template does not introduce a significant amount of additional noise, and that the signal-to-noise ratio
is dominated by the WMAP9 data\footnote{This is consistent with the reduced sky fraction employed for the \planck\ LFI 70 GHz channel in the 2015 release as opposed to WMAP (47\% vs.\ 73\%) and the fact that \planck\ uses only one channel as apposed to WMAP's three.}.

It is also interesting to note how, for a given low-$\ell$ dataset, the $\tau$ estimate is remarkably stable with respect to the addition of 
high-$\ell$ and/or lensing data. For example, adding the \plikTT\ data we get:
\begin{subequations}
\begin{align}
&\tau = 0.091^{+0.011} _{-0.013}, \quad \textrm{\wlowP\ + \plikTT},\\
&\tau = 0.066\pm 0.013, \quad \textrm{\whfilowP\ + \plikTT}. \label{eq:tau_w353_plikTT}\\
&\tau = 0.068^{+0.011}_{-0.013}, \quad \textrm{\lfiwhfilowP\ + \plikTT } \label{eq:tau_lfiw353_plikTT},
\end{align}
\emph{i.e}, sub-sigma shift with respect to the corresponding low-$\ell$-only values, see Eqs. (\ref{eq:tau_WP9}), (\ref{eq:tau_WP353}), (\ref{eq:tau_WPLFI}). This also happens when using \wmapTT\ and/or adding \lensing\, but not when using \lowP\ which, in combination with \plikTT\  data gives 
\citep{Ade:2015xua}:
\begin{equation}
\tau =  0.078 \pm 0.019 \quad \textrm{\lowP\ + \plikTT}\, \label{eq:tau_lowPplTT}.
\end{equation}
\end{subequations}
This value is significantly larger than both the low-ell-only estimates (\ref{eq:tau_lowP}) and (\ref{eq:tau_WPLFI}), and than the estimates (\ref{eq:tau_w353_plikTT}) and
(\ref{eq:tau_lfiw353_plikTT}) obtained including high-ell's. This shift can be ascribed to the weaker constraining power of the \planck\ 70GHz-based low-$\ell$ polarization dataset, 
giving a larger statistical weight to the high $\ell$'s relative to the low $\ell$'s. In fact,
when the larger uncertainty is taken into account, the \emph{relative} shift between (\ref{eq:tau_lowPplTT}) and (\ref{eq:tau_lowP})
is found to be quite small $(\sim 0.4 \sigma)$. When the lensing information is added, $\tau$ shifts back
to lower values \citep{Ade:2015xua}:
\begin{align}
&\tau =  0.066 \pm 0.016 \quad \textrm{\lowP\ + \plikTT} \nonumber \\ 
& \eightquad
\textrm{+ \lensing}\, .
\end{align}
Finally, completely disregarding the low-$\ell$ information (including temperature) and using only the high-$\ell$ and lensing data from \planck\ yields:
\begin{align}
\tau = 0.066^{+0.025}_{-0.026} \quad \textrm{\plikTT\ + \lensing},
\end{align}
in excellent agreement with the results obtained with \whfilowP, and still in tension with the \wlowP\ result, although 
with a lower significance ($\sim 1\sigma$) due to the larger uncertainty.

We now focus our attention on the amplitude of primordial scalar perturbations, $\log(10^{10} A_s)$ and on
the combination $\clampp$ that effectively governs the overall amplitude of CMB anisotropies 
at scales below the size of the horizon at reionization, corresponding to a multipole $\ell_\mathrm{reio} \sim 10$. Looking first at the effect of the
\planck\ 353 GHz-based cleaning, it can be noticed that it gives a systematically lower value for $\log(10^{10} A_s)$
than the WMAP dust template. For example, from the low-$\ell$ data together with the \plikTT\ + \lensing\ dataset:
\begin{subequations}
\begin{align}
&\log(10^{10} A_s) = 3.095^{+0.020}_{-0.022}\, ,  \qquad\textrm{\wlowP\ + }  \nonumber \\
& \nquad{6.5} \textrm{+ \plikTT\ + \lensing\,;}\\[0.2cm]
&\log(10^{10} A_s) = 3.055^{+0.023}_{-0.022}\, ,  \qquad\textrm{\whfilowP\ + }  \nonumber \\
& \nquad{6.5} \textrm{+ \plikTT\ + \lensing\,;}\\[0.2cm]
&\log(10^{10} A_s) = 3.060 \pm 0.020\, ,  \qquad\textrm{\lfiwhfilowP\ + }  \nonumber \\
& \nquad{6.5} \textrm{+ \plikTT\ + \lensing\,.}
\end{align}
\end{subequations}
For comparison $\log(10^{10} A_s) = 3.062 \pm {0.029}$ from the analysis in \citet{Ade:2015xua}, using
\lowP\ + \plikTT\ + \lensing.

This dependence of $\log(10^{10} A_s)$ estimates from the cleaning procedure should not come as a surprise, as it is simply an 
effect of the degeneracy between $A_s$ and $\tau$: given a determination of $\clampp$
from observations at $\ell > \ell_\mathrm{reio}$, a larger value of $\tau$, such as the one preferred by the original WMAP9 polarization data,
requires in turn a larger $A_s$ to keep $\clampp$ constant. This is particularly evident when
high-$\ell$ data are included, as this leads to a more precise determination of $\clampp$. This reasoning is supported by the fact
that derived values of $\calAs \equiv \clamp$ are instead quite consistent with respect to the choice of the cleaning procedure:
\begin{subequations}
\begin{align}
&\calAs = 1.866 \pm 0.012\, ,  \qquad\textrm{\wlowP\ + }  \nonumber \\
& \nquad{6.5} \textrm{+ \plikTT\ + \lensing\,;}\\[0.2cm]
&\calAs = 1.876\pm 0.012\, ,  \qquad\textrm{\whfilowP\ + }  \nonumber \\
& \nquad{6.5} \textrm{+ \plikTT\ + \lensing\,;}\\[0.2cm]
&\calAs = 1.874 \pm 0.012\, ,  \qquad\textrm{\lfiwhfilowP\ + }  \nonumber \\
& \nquad{6.5} \textrm{+ \plikTT\ + \lensing\,.}
\end{align}
\end{subequations}

Focusing instead on the impact of the high-$\ell$ dataset on measurements of the scalar amplitude, we find that 
low-$\ell$ data alone prefer lower values of the amplitude than the high-$\ell$ data. This is a well-known
reflection of the so-called ``low-$\ell$ anomaly'' \citep{Bennett:2003bz,Ade:2013zuv,Ade:2015xua,Aghanim:2016sns}.
We note that this low- vs. high-$\ell$ tension
is not alleviated by the use of the \planck\ 353 GHz for cleaning (pointing to the fact that it is not related
to the degeneracy with $\tau$), nor does it depend on the particular high-$\ell$ data considered. 
In addition to this tension, there are also
small but noticeable differences among the $\calAs$ values inferred using different high-$\ell$ datasets. 
In general, the high-$\ell$ TT data from WMAP9
lead to lower estimates of the amplitude than the corresponding \planck\ data, while
estimates from \lensing\ seem to sit somehow in the middle, as its inclusion brings the results from the two
TT datasets closer to each other. Fixing for definiteness the low-$\ell$'s in polarization to be those from the 
noise-weighted WMAP/\planck\ maps,
we get:
\begin{subequations}
\begin{align}
&\calAs = 1.702^{+0.085}_{-0.099}\, ; \qquad \textrm{\lfiwhfilowP}, \label{eq:clamp_lowell} \\[0.2cm]
&\calAs = 1.834 \pm 0.031\,; \quad \textrm{+ \wmapTT}, \\[0.2cm]
&\calAs  = 1.858 \pm 0.017; \quad \textrm{+ \wmapTT\ } \nonumber \\
&\nquad{11} \textrm{+ \lensing}, \\[0.2cm]
&\calAs  = 1.883 \pm 0.014; \quad \textrm{+ \plikTT},\\[0.2cm]
&\calAs  = 1.874^{+0.011}_{-0.012}; \quad \textrm{+ \plikTT\ + \lensing},
\end{align}
while using the \planck\ high-$\ell$ and lensing data alone we get
\begin{align}
&\calAs = 1.875^{+0.015}_{-0.016}; \quad \textrm{\plikTT + \lensing}.
\end{align}
\end{subequations}
Comparing this last value with (\ref{eq:clamp_lowell}), we can quantify the tension between high- and low-$\ell$'s
determinations of the amplitude to be at the $2\sigma$ level.

\begin{figure}
\begin{center}
\includegraphics[width=0.6\linewidth,keepaspectratio]{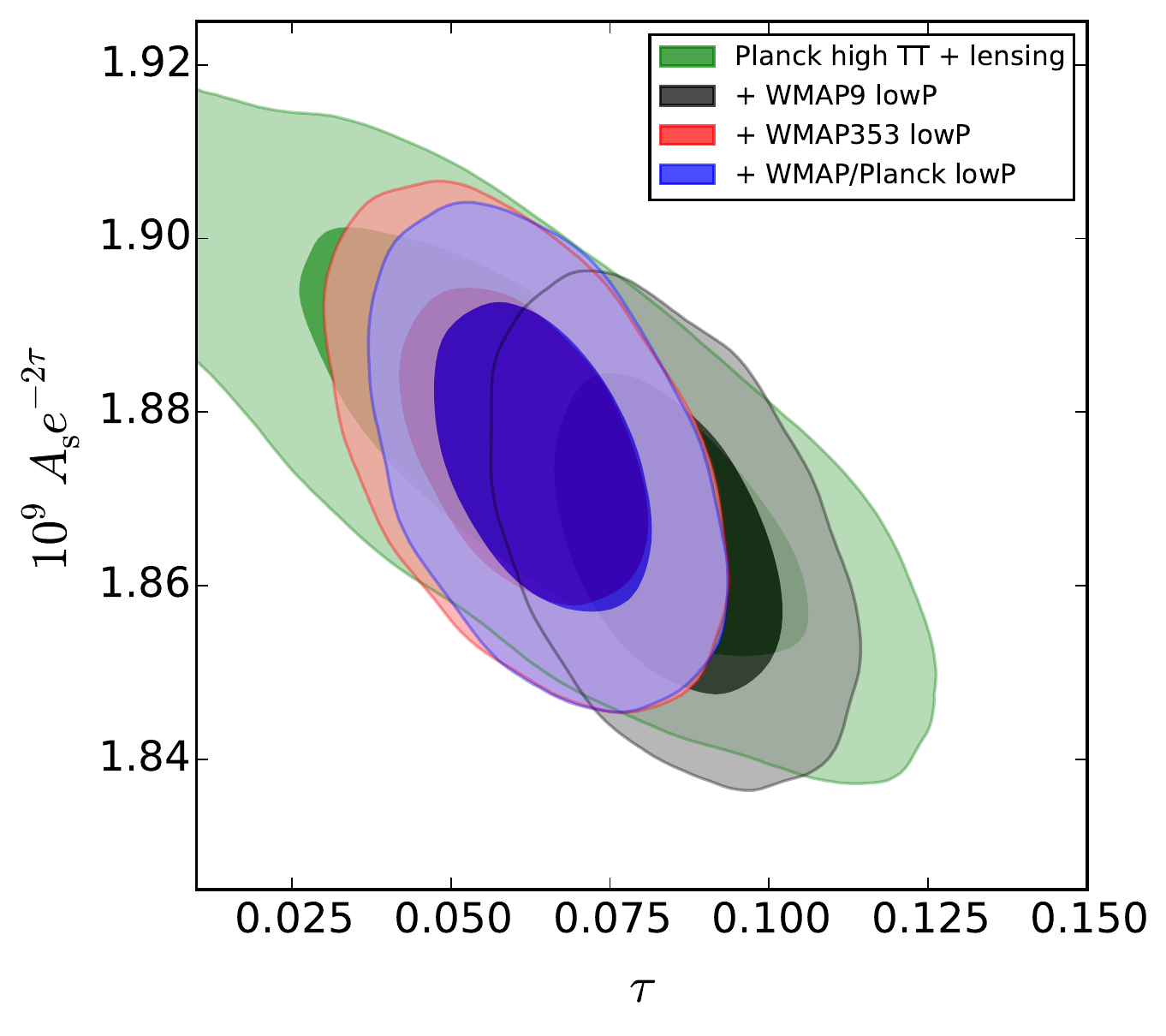}%
\caption{\label{fig:post2D_low} Impact of the choice of the low-$\ell$ dataset on estimates of $\tau$ and $\clamp$. Light (dark) shaded
contours show 68\% (95\%) credible intervals in the $(\tau,\,\clamp)$ plane. Green contours are obtained
using the \planck\ high-$\ell$ TT and lensing data only, \emph{i.e.}, without using information from $\ell < 30$.
The black, red and blue regions show the effect of adding the low-$\ell$ datasets indicated in the legend
(as emphasized in the text, the low-$\ell$ temperature data always comes from \planck).}
\end{center}
\end{figure}

\begin{figure*}
\begin{center}
\includegraphics[width=0.49\linewidth,keepaspectratio]{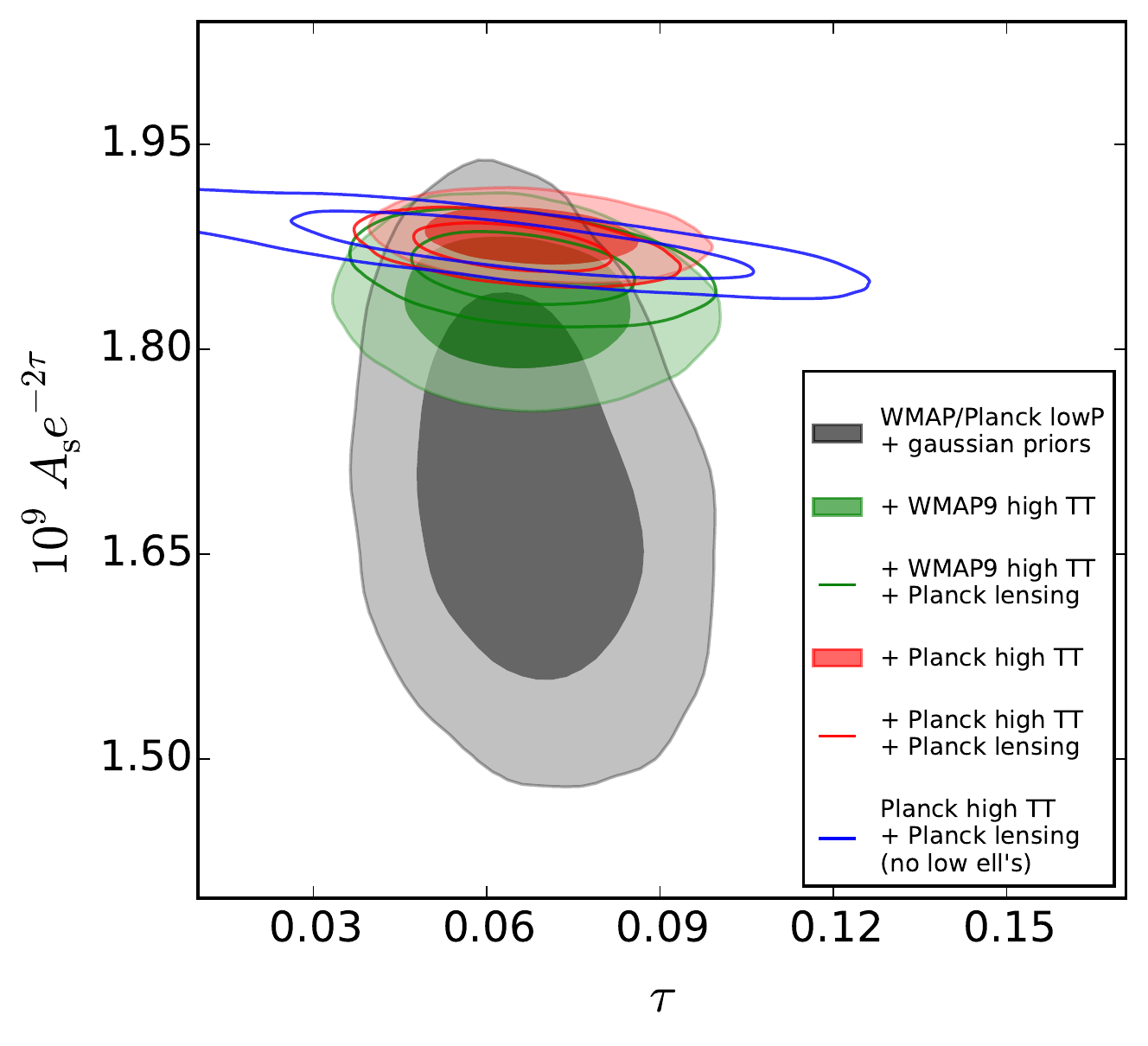}%
\includegraphics[width=0.49\linewidth,keepaspectratio]{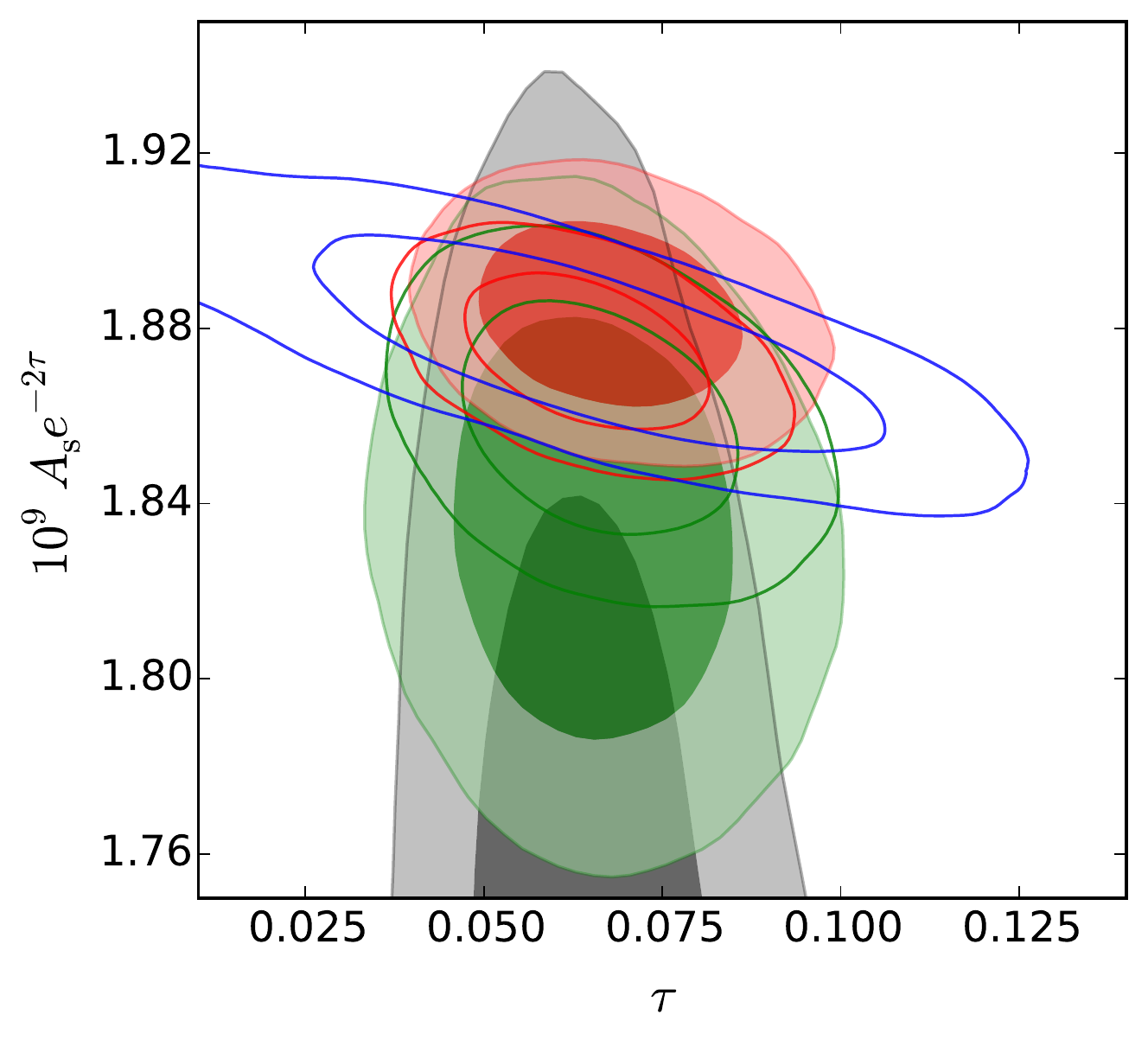}
\caption{\label{fig:post2D_hi} \textit{Left panel:} Impact of the choice of the high-$\ell$ dataset on estimates of $\tau$ and $\clamp$. 
We show 68\% (95\%) credible regions in the $(\tau,\,\clamp)$ plane. The low-$\ell$ dataset
is always provided by the \planck\ Commander map in temperature and by the combined W353/LFI70 maps
in polarization. Grey contours are obtained from this dataset in combination with gaussian priors on the ``high-$\ell$'' parameters.
The green and red curves combine the low-$\ell$ data with WMAP9 or \planck\ high-$\ell$ TT data, respectively;
the unfilled curves also use \planck\ lensing. Finally, the empty blue contours are obtained from 
\planck\ high-$\ell$ TT and lensing data, without information coming from $\ell < 30$. 
\textit{Right panel}: The same as the left panel, but zoomed to make the effect of the high-$\ell$ datasets on the determination of $\clamp$ more
evident.}
\end{center}
\end{figure*}

We summarize our findings on $\calAs$ and $\tau$ in Figures \ref{fig:post2D_low} and \ref{fig:post2D_hi}.
In particular, in figure \ref{fig:post2D_low} we show the impact of the choice of the low-$\ell$ dataset
on the determination of these parameters, while in figure \ref{fig:post2D_hi} we focus on the effect
of the high-$\ell$ dataset.

\section{Conclusions}
\label{conclusions}

In this paper we have studied the impact of low-$\ell$ and high-$\ell$ CMB data on the estimation
of cosmological parameters. At first, we have focused on the choice of the large-scale polarisation data,
finding that cleaning the WMAP polarization data with the \planck\ 353 GHz dust template yields a significantly lower value of $\tau$ with respect to the same 
data cleaned using the WMAP9 dust model; this also affects estimates of the scalar amplitude $A_s$ through its degeneracy with $\tau$.
Concerning Galactic synchrotron and thermal dust polarized emissions, we find spectral indexes in agreement with previous works and robust evidence of the gray body shape 
of the dust emission spectrum.
Combining the WMAP and \planck\ 70 GHz polarization data, and cleaning with the \planck\ dust template,
we find $\tau = 0.066 ^{+0.012}_{-0.013}$; this value is very stable against the inclusion of high-$\ell$ data (including \planck\ lensing) 
and is in remarkable agreement with the \planck\ low-$\ell$ only estimate $\tau = 0.067 \pm 0.023$, as well as being consistent
with the more recent value $\tau = 0.055 \pm 0.009$ obtained by the \planck\ collaboration using the large-scale polarisation data
from the high-frequency instrument \citep{Aghanim:2016yuo}. We also find a remarkable consistency with the estimate
$\tau = 0.066 ^{+0.025}_{-0.026}$ obtained disregarding the low-$\ell$ information and using \plikTT\ and \lensing\ data, that independently measure 
$A_s e^{-2\tau}$ and $A_s$, and thus allow an indirect determination of the optical depth.\footnote{See Ref.  \cite{Gnedin:2000uj} for a theoretical model predicting a value
of $\tau$ in agreement with those discussed in this paper.} Finally, we note that
even if $\tau$ and $\log(10^{10} A_s)$ are the parameters which are more influenced by the cleaning of the polarization maps, other base parameters 
are also affected, albeit to a lesser extent. This is especially evident when the \plikTT\ and \lensing\ are used jointly with the low-$\ell$ polarization data: while $\tau$ and $\log(10^{10} A_s)$ shift by roughly $1.2\sigma$ between the \planck\ 353 GHz and WMAP dust model cleanings, the estimates of the other base parameter shift by an amount between $\sim 0.4 \sigma$ (for $\theta$) and $~0.8 \sigma$ (for $\Omega_c h^2$).
More detailed information can be found in table \ref{tab:shifts}, where we report the parameter estimates obtained by combining different
low-$\ell$ polarization datasets (including the case where none at all is employed) with \plikTT\ and \lensing, as well as the corresponding parameter shifts with
respect to the \wlowP\ + \plikTT\ + \lensing\ case.
The value of the Hubble constant that is obtained from the \wlowP\ + \plikTT\ +\lensing\ combination is also $\sim 1\sigma$ lower than the corresponding
value from datasets cleaned with the \planck\ dust templates. In general, we find that values of the Hubble constant obtained using polarization datasets
cleaned with the \planck\ 353 GHz channel are systematically higher than the corresponding value obtained with \wlowP, as can be seen by looking at the relevant panel of figure \ref{fig:whisk_derived} and comparing the black whiskers on one side and the red and green whiskers on the other. This contributes 
to increase the tension with local measurements of $H_0$ \cite{Riess:2016jrr}. We finally remark that for all parameters, the estimates from the \wlowP\ + \plikTT\ +\lensing\ 
dataset combination are in slight tension with those obtained from \plikTT\ + \lensing. 

\begin{landscape}
\begin{table}
\centering
\caption{Parameter estimates from \plikTT + \lensing\, either in combination with the low-$\ell$ datasets
considered in our analysis (second to fourth column) or alone (fifth column). In the former cases, use of \plancklowT\ is understood.
The last three columns report 
the shifts (in units of $\sigma$) in the parameters with respect to \wlowP\ + \plikTT\ + \lensing.
}
\label{tab:shifts}
\begin{tabular}{cccccccc}
\hline
& \multicolumn{4}{c}{Low-$\ell$ dataset} & &  &\\
& \multicolumn{4}{c}{(all are + \plikTT + \lensing)} & &  &\\[0.2cm]
Parameter & [1] \wlowP  & [2] \whfilowP & [3] \lfiwhfilowP & [4] no low-$\ell$ data &$\frac{[2] - [1]}{\sqrt{\sigma^2_{[2]}+\sigma^2_{[1]}}}$ & $\frac{[3] - [1]}{\sqrt{\sigma^2_{[3]}+\sigma^2_{[1]}}}$ &$\frac{[4] - [1]}{\sqrt{\sigma^2_{[4]}+\sigma^2_{[1]}}}$  \\
\hline
\hline
$\Omega_b h^2$ & $0.02238\pm0.00022$ & $0.02222\pm0.00021$ & $0.02224\pm0.00021$ & $0.02223^{+0.00026}_{-0.00027}$ & $-0.53$ & $-0.45$ & $-0.43$ \\[0.1cm] 
 $\Omega_c h^2$ & $0.1169\pm0.0017$ & $0.1190\pm0.0017$ & $0.1187\pm0.0017$   & $0.1189\pm0.0027$& $+0.87$ & $+0.77$ & $+0.65$\\[0.1cm] 
 $100 \theta$ & $1.04127\pm0.00043$ & $1.04096\pm0.00044$ & $1.04101\pm0.00043$  & $1.04101^{+0.00052}_{-0.00053}$ &$-0.50$& $-0.43$ & $-0.39$\\[0.1cm] 
 $\tau$ & $0.084^{+0.011}_{-0.012}$ & $0.062\pm0.012$ & $0.065^{+0.010}_{-0.012}$ & $0.066^{+0.025}_{-0.026}$ & $-1.3$& $-1.2$ & $-0.64$ \\[0.1cm] 
 $\log[10^{10}A_s]$ & $3.095^{+0.020}_{-0.022}$ & $3.055\pm0.023$ & $3.060\pm0.020$ & $3.064\pm 0.045$&$-1.3$& $-1.2$ & $-0.61$\\[0.1cm] 
 $n_s$ & $0.9723\pm0.0052$ & $0.9665\pm0.0053$ & $0.9673\pm0.0051$  & $0.9657\pm 0.0078$ & $-0.78$ & $-0.68$ & $-0.71$\\[0.1cm] 
\hline 
$10^9 A_s \exp[-2\tau]$ & $1.866\pm0.012$ & $1.876\pm0.012$ & $1.874\pm0.011$ &$1.875^{+0.015}_{-0.016}$ &$+0.62$& $+0.54$ &$+0.50$ \\[0.1cm] 
 $H_0$ [km/s/Mpc] & $68.59\pm0.77$ & $67.61\pm0.80$ & $67.74\pm0.76$ & $67.7^{+1.2}_{-1.3}$&$-0.89$& $-0.79$ & $-0.64$\\[0.1cm] 
 $\Omega_m$ & $0.2975\pm0.0098$ & $0.310\pm0.011$ & $0.309\pm0.010$ & $0.310\pm0.017$ & $+0.89$ & $+0.79$ & $+0.66$\\[0.1cm] 
\hline
\hline
\end{tabular}
\end{table}
\end{landscape}

As for the role of the high-$\ell$ dataset, we confirm the existence of the so-called ``low-$\ell$ anomaly'', i.e., 
the fact that the value of the amplitude parameter $\calAs$ estimated by the low-$\ell$ data is lower, at the $\sim 2 \sigma$ level,
than the one measured from the high-$\ell$ data. This anomaly is not affected by the cleaning procedure used on the large-scale polarization. 
The low-ell anomaly has been found to drive the shifts in parameter values obtained from different multipole ranges
\citep{Addison:2015wyg,Aghanim:2016sns}, so our findings imply that these shifts are not affected by the choice of low-ell polarization data.
On the other hand, values of the amplitude $\calAs$ are fairly consistent between the WMAP and \planck\ high-$\ell$ TT datasets, as well as with \lensing.

\acknowledgments

This paper is based on observations obtained with the satellite \planck\ (\url{http://www.esa.int/Planck}),
an ESA science mission with instruments and contributions directly funded by ESA Member
States, NASA, and Canada.
We acknowledge support from ASI through ASI/INAF Agreement 2014-
024-R.1 for the Planck LFI Activity of Phase E2.
We acknowledge the use of computing facilities at NERSC (USA). 
We acknowledge use of the HEALPix \citep{Gorski:2004by} software and analysis package for deriving the results in this paper. 
We acknowledge the use of the Legacy Archive for Microwave Background Data Analysis (LAMBDA). Support for LAMBDA is provided by the NASA Office of Space Science. MG acknowledges support by Katherine Freese through a grant from the Swedish Research Council (Contract No. 638-2013-8993).
We thank Loris Colombo, Luca Pagano, Bruce Partridge and Douglas Scott for useful discussion and comments on the manuscript.




\begin{thebibliography}{99}

\bibitem{Fixsen:2009ug} 
  D.~J.~Fixsen,
  \textit{The Temperature of the Cosmic Microwave Background},
  \textit{Astrophys.\ J.}\  {\bf 707}  (2009) 916
  [arXiv:0911.1955].
  
\bibitem{Hu:1997hv} 
  W.~Hu and M.~J.~White,
  \textit{A CMB polarization primer},
  \textit{New Astron.} \  {\bf 2} (1997) 323 
  [astro-ph/9706147].
  
\bibitem{Gunn:1965hd} 
  J.~E.~Gunn and B.~A.~Peterson,
  \textit{On the Density of Neutral Hydrogen in Intergalactic Space},
  \textit{Astrophys.\ J.}\  {\bf 142}  (1965) 1633.
  
  
\bibitem{Fan:2005es} 
  X.~H.~Fan {\it et al.},
  \textit{Constraining the evolution of the ionizing background and the epoch of reionization with z~6 quasars. 2. a sample of 19 quasars}
  \textit{Astron.\ J.}\  {\bf 132} (2006) 117
  [astro-ph/0512082].
  
\bibitem[{Bolton et al.}(2011)]{Bolton:2011vb} 
  J.~S.~Bolton, M.~G.~Haehnelt, S.~J.~Warren, P.~C.~Hewett, D.~J.~Mortlock, B.~P.~Venemans, R.~G.~McMahon and C.~Simpson,
  \textit{How neutral is the intergalactic medium surrounding the redshift z=7.085 quasar ULAS J1120+0641?},
  \textit{Mon.\ Not.\ Roy.\ Astron.\ Soc.}\  {\bf 416} (2011) L70
  [arXiv:1106.6089].
  
\bibitem[{Chornock et al}(2013)]{Chornock:2013una} 
  R.~Chornock, E.~Berger, D.~B.~Fox, R.~Lunnan, M.~R.~Drout, W.~F.~Fong, T.~Laskar and K.~C.~Roth,
  \textit{GRB 130606A as a Probe of the Intergalactic Medium and the Interstellar Medium in a Star-forming Galaxy in the First Gyr After the Big Bang},
  \textit{Astrophys.\ J.} \  {\bf 774} (2013) 26 
  [arXiv:1306.3949]. 
  
\bibitem[{McGreer, Mesinger and D'Odorico}(2015)]{McGreer:2014qwa} 
  I.~McGreer, A.~Mesinger and V.~D'Odorico,
  \textit{Model-independent evidence in favour of an end to reionization by $z \approx$ 6},
  \textit{Mon.\ Not.\ Roy.\ Astron.\ Soc.}\  {\bf 447} (2015) 499
  [arXiv:1411.5375].

\bibitem[{\planck\ 2015 results XIII}()]{Ade:2015xua} 
   Planck Collaboration,
     \textit{Planck 2015 results. XIII. Cosmological parameters,}
   \textit{Astron.\ \& Astrophys.}\, \textbf{594} (2016) A13
    [arXiv:1502.01589].
  
\bibitem[{Mortonson and Hu}(2008)]{Mortonson:2007hq} 
  M.~J.~Mortonson and W.~Hu,
  \textit{Model-independent constraints on reionization from large-scale CMB polarization},
  \textit{Astrophys.\ J.}\  {\bf 672} (2008) 737
  [arXiv:0705.1132].
  
\bibitem[{Pandolfi et al.}(2011)]{Pandolfi:2011kz} 
  S.~Pandolfi, A.~Ferrara, T.~R.~Choudhury, A.~Melchiorri and S.~Mitra,
  \textit{Data-constrained reionization and its effects on cosmological parameters},
  \textit{Phys.\ Rev.\ D} {\bf 84} (2011) 123522
  [arXiv:1111.3570].
  
    \bibitem[Trombetti and Burigana(2012)]{Trombetti:2012}
  T.~Trombetti and C.~Burigana,
  \textit{Imprints on CMB Angular Power Spectrum Modes from Cosmological Reionization},
  \textit{J. Mod. Phys.} \textbf{3} (2012) 1918.

\bibitem[{Douspis et al.}(2015)]{Douspis:2015nca} 
  M.~Douspis, N.~Aghanim, S.~Ili\'c and M.~Langer,
  \textit{A new parameterization of the reionisation history},
  \textit{Astron.\ Astrophys.}\  {\bf 580} (2015) L4
  [arXiv:1509.02785].
  
\bibitem[{Oldengott, Boriero and Schwarz}(2016)]{Oldengott:2016yjc} 
  I.~M.~Oldengott, D.~Boriero and D.~J.~Schwarz,
  \textit{Reionization and dark matter decay},
  \textit{JCAP} {\bf 08} (2016) 054
  [arXiv:1605.03928].
  
\bibitem[{Planck Collaboration Int. XLVII}(2016)]{Adam:2016hgk} 
  Planck Collaboration, 
  \textit{Planck intermediate results. XLVII. Planck constraints on reionization history},
\textit{Astron.\ Astrophys.} in press
  [arXiv:1605.03507].
  
\bibitem[{Heinrich, Miranda and Hu}(2016)]{Heinrich:2016ojb} 
  C.~H.~Heinrich, V.~Miranda and W.~Hu,
  \textit{Complete Reionization Constraints from Planck 2015 Polarization},
  arXiv:1609.04788.
  
\bibitem[{Hinshaw et al.}(2013)]{Hinshaw:2012aka} 
  G.~Hinshaw {\it et al.} [WMAP Collaboration],
  \textit{Nine-Year Wilkinson Microwave Anisotropy Probe (WMAP) Observations: Cosmological Parameter Results},
  \textit{Astrophys.\ J.\ Suppl}.\  {\bf 208} (2013) 19
  [arXiv:1212.5226].
  
\bibitem[{Planck Collaboration 2015 I}(2016)]{Adam:2015rua} 
  Planck Collaboration,
  \textit{Planck 2015 results. I. Overview of products and scientific results,}
  \textit{Astron.\ Astrophys.}\  {\bf 594} (2016) A1
  [arXiv:1502.01582].
  
  
\bibitem{Ade:2013kta} 
   Planck Collaboration,
  \textit{Planck 2013 results. XV. CMB power spectra and likelihood},
  \textit{Astron.\ Astrophys.}\  {\bf 571} (2014) A15
  [arXiv:1303.5075].
  
\bibitem[{Planck Collaboration 2015 XI}(2016)]{Aghanim:2015xee} 
   Planck Collaboration,
    \textit{Planck 2015 results. XI. CMB power spectra, likelihoods, and robustness of parameters},
    \textit{Astron.\ \& Astrophys.}\  \textbf{594} (2016) A11
 [arXiv:1507.02704].
  
\bibitem[{Planck Collaboration Int. XLVI}(2016)]{Aghanim:2016yuo} 
  Planck Collaboration,
  \textit{Planck intermediate results. XLVI. Reduction of large-scale systematic effects in HFI polarization maps and estimation of the reionization optical depth},
  arXiv:1605.02985 [astro-ph.CO].

\bibitem[{Addison et al.}(2015)]{Addison:2015wyg} 
  G.~E.~Addison, Y.~Huang, D.~J.~Watts, C.~L.~Bennett, M.~Halpern, G.~Hinshaw and J.~L.~Weiland,
  \textit{Quantifying discordance in the 2015 Planck CMB spectrum},
  \textit{Astrophys.\ J.}\  {\bf 818}, (2016) 132
  [arXiv:1511.00055].

\bibitem[Planck Collaboration Int. LI(2016)]{Aghanim:2016sns} 
  Planck Collaboration,
  \textit{Planck 2016 intermediate results. LI. Features in the cosmic microwave background temperature power spectrum and shifts in cosmological parameters},
  arXiv:1608.02487 [astro-ph.CO].
  
\bibitem[{Couchot et al.}(2015)]{Couchot:2015eea} 
  F.~Couchot, S.~Henrot-Versill\'e, O.~Perdereau, S.~Plaszczynski, B.~R.~d'Orfeuil and M.~Tristram,
  \textit{Relieving tensions related to the lensing of CMB temperature power spectra,}
  arXiv:1510.07600 [astro-ph.CO].
  
\bibitem{Gorski:2004by} 
  K.~M.~Gorski, E.~Hivon, A.~J.~Banday, B.~D.~Wandelt, F.~K.~Hansen, M.~Reinecke and M.~Bartelman,
  \textit{HEALPix - A Framework for high resolution discretization, and fast analysis of data distributed on the sphere},
  \textit{Astrophys.\ J.}\  {\bf 622} (2005) 759
  [astro-ph/0409513].

 
\bibitem[{Bennett et al.}(2013)]{Bennett:2012zja} 
  C.~L.~Bennett {\it et al.} [WMAP Collaboration],
  \textit{Nine-Year Wilkinson Microwave Anisotropy Probe (WMAP) Observations: Final Maps and Results},
  \textit{Astrophys.\ J.\ Suppl.}\  {\bf 208} (2013) 20
  [arXiv:1212.5225].

\bibitem[{Haslam et al.}(1982)]{Haslam:1982zz} 
  C.~G.~T.~Haslam, C.~J.~Salter, H.~Stoffel and W.~E.~Wilson,
  \textit{A 408 MHz all-sky continuum survey. II. The atlas of contour maps},
  \textit{Astron.\ Astrophys.\ Suppl.\ Ser.}\  {\bf 47}, (1982) 1.
 
\bibitem[{Planck Collaboration 2015 IX}(2016)]{Adam:2015tpy} 
  Planck Collaboration,
  \textit{Planck 2015 results. IX. Diffuse component separation: CMB maps},
  \textit{Astron.\ Astrophys.}\  {\bf 594} (2016) A9
  [arXiv:1502.05956].

\bibitem[{Planck Collaboration 2015 X}(2016)]{Adam:2015wua} 
  Planck Collaboration,
  \textit{Planck 2015 results. X. Diffuse component separation: Foreground maps},
  \textit{Astron.\ Astrophys.}\  {\bf 594} (2016) A10)
  [arXiv:1502.01588].
  
\bibitem[Eriksen et al.(2007)]{Eriksen:2007mx} 
  H.~K.~Eriksen, J.~B.~Jewell, C.~Dickinson, A.~J.~Banday, K.~M.~Gorski and C.~R.~Lawrence,
  \textit{Joint Bayesian component separation and CMB power spectrum estimation},
  \textit{Astrophys.\ J.}\  {\bf 676} (2008) 10
  [arXiv:0709.1058].
  
\bibitem[Eriksen et al.(2004)]{Eriksen:2004ss} 
  H.~K.~Eriksen {\it et al.},
  \textit{Power spectrum estimation from high-resolution maps by Gibbs sampling,}
  \textit{Astrophys.\ J.\ Suppl.}\  {\bf 155} (2004) 227
  [astro-ph/0407028].
  
\bibitem[Tegmark and de Oliveira-Costa(2001)]{Tegmark:2001zv}
  M.~Tegmark and A.~de Oliveira-Costa,
  \textit{How to measure CMB polarization power spectra without losing information},
  \textit{Phys.\ Rev.\ D} {\bf 64} (2001) 063001
  [astro-ph/0012120].

\bibitem[Page et al.(2007)]{Page:2006hz} 
  L.~Page {\it et al.}  [WMAP Collaboration],
  \textit{Three year Wilkinson Microwave Anisotropy Probe (WMAP) observations: polarization analysis},
  \textit{Astrophys.\ J.\ Suppl.}\  {\bf 170} (2007) 335
  [astro-ph/0603450].
  
\bibitem[{Planck Collaboration Int. XXX}(2016)]{Adam:2014bub} 
  Planck Collaboration,
  \textit{Planck intermediate results. XXX. The angular power spectrum of polarized dust emission at intermediate and high Galactic latitudes},
  \textit{Astron.\ Astrophys.}\  {\bf 586} (2016) A133
  [arXiv:1409.5738].
  
\bibitem[{Planck Collaboration 2015 XV}(2016)]{Ade:2015zua} 
   Planck Collaboration,
   \textit{Planck 2015 results. XV. Gravitational lensing},
    \textit{Astron.\ \& Astrophys.}\ \textbf{594} (2016) A15.
  [arXiv:1502.01591].
  
\bibitem[{Lewis and Bridle}(2002)]{Lewis:2002ah} 
  A.~Lewis and S.~Bridle,
  \textit{Cosmological parameters from CMB and other data: A Monte Carlo approach},
  \textit{Phys.\ Rev.\ D} {\bf 66} (2002) 103511.
  [astro-ph/0205436].

\bibitem[{Lewis et al.}(1999)]{Lewis:1999bs} 
  A.~Lewis, A.~Challinor and A.~Lasenby,
  \textit{Efficient computation of CMB anisotropies in closed FRW models},
  \textit{Astrophys.\ J.}\  {\bf 538} (2000) 473
  [astro-ph/9911177].


\bibitem[{Bennett et al.}(2003)]{Bennett:2003bz} 
  C.~L.~Bennett {\it et al.} [WMAP Collaboration],
  \textit{First year Wilkinson Microwave Anisotropy Probe (WMAP) observations: Preliminary maps and basic results},
  \textit{Astrophys.\ J.\ Suppl.}\  {\bf 148} (2003) 1
  [astro-ph/0302207].
  
\bibitem[{Planck 2013 XVI}(2014)]{Ade:2013zuv} 
  Planck Collaboration,
  \textit{Planck 2013 results. XVI. Cosmological parameters},
  \textit{Astron.\ Astrophys.}\  {\bf 571} (2014) A16
[arXiv:1303.5076].
  
\bibitem[Gnedin(2000)]{Gnedin:2000uj} 
  N.~Y.~Gnedin,
  \textit{Effect of reionization on the structure formation in the universe},
  \textit{Astrophys.\ J.}\  {\bf 542} (2000)  535
  [astro-ph/0002151].

\bibitem{Riess:2016jrr} 
  A.~G.~Riess {\it et al.},
\textit{A 2.4\% Determination of the Local Value of the Hubble Constant},
\textit{Astrophys.\ J.}\  {\bf 826} (2016) 56
  [arXiv:1604.01424].
    
 
  
%
%





\end{thebibliography}
\end{document}